\documentclass[12pt]{article}
\usepackage{graphics}
\usepackage{subfigure}
\usepackage{epsfig}
\begin{document}
\title{Hadron transverse momentum distributions and TMD studies}
\author{Jean-Francois Rajotte\thanks{E-mail: jeanfrancois.rajotte@cern.ch} \\ 
on behalf of the COMPASS Collaboration \\
 Ludwig-Maximilians Universiteat, Fakultaet fuer Physik, \\
 D-80799 Munich, Germany\\
}
\date{}
\maketitle

\abstract{
Charged hadron differential distributions from muon-induced deep inelastic scattering, DIS, on a $^6$LiD target are presented as function of the 
DIS variables $x_{Bj}$, $Q^2$, $W^2$ and the hadron variables $p_T$ and $z$.  
They can be used as benchmark to verify the reliability of theoretical model.  
The $p_T^2$ distributions are fitted with a Gaussian function at different kinematic intervals.  
With a Gaussian ansatz for the transverse momentum dependent parton distributions, TMDs, 
the intrinsic transverse momentum of the partons is extracted.
} %end of abstract
\maketitle
\section{Introduction}
\label{s_intro}
\begin{figure}[t]
      \begin{center}
            \resizebox{0.7\columnwidth}{!}{%
            \includegraphics{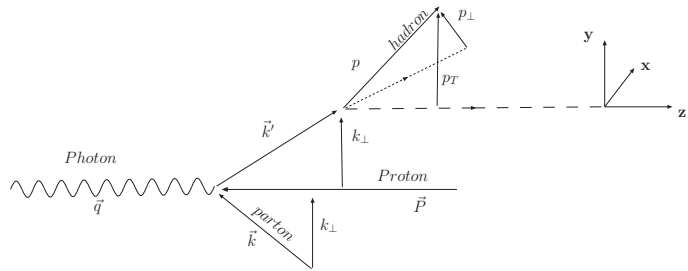} }
      \end{center}
\caption{Kinematic variables of virtual photon interaction with a parton with intrinsic transverse momentum 
%$\textbf{k}_{\perp}$ 
$k_{\perp}$ and its hadronization.  The transverse momentum of the observed hadron, $p_T$, is defined with respect to the virtual photon.}
\label{f_intrinsicKin}       % Give a unique label
\end{figure}
Semi Inclusive Deep Inelastic Scattering (SIDIS) reactions provide much information about the structure of the nucleon and the hadronization of partons.  
%Already in the 1970s (see \cite{Cahn:1978se} and reference therein), 
%Already in the 1970s, SIDIS was connected to the intrinsic transverse momentum of the partons as shown in figure \ref{f_intrinsicKin}.  
Parton intrinsic transverse momentum affects the transverse momentum of the produced hadron.  
The kinematic variables of the produced hadron are shown in figure \ref{f_intrinsicKin}, 
where the transverse momentum is determined w.r.t. the virtual photon.  
The transverse momentum distributions from unpolarized scattering 
are often the first verification of the reliability of a theoretical model.
%There are several effects which contribute to the shape of the $p_{T}^2$ distributions, 
The shape of the $p_{T}^2$ distributions depends on many effects, 
%e.g. TMDs, fragmentation functions (FFs) and gluon radiation.  
e.g. contribution from intrinsic transverse momentum $k_{\perp}$ to the TMDs 
and transverse momentum of hadron $p_{\perp}$ to the fragmentation functions (FFs) and gluon radiation.  
These effects depend on kinematic variables such as the Bjorken variable $x_{Bj}$, 
the invariant mass squared $W^2$ of the hadronic system, 
the negative 4-momentum squared $Q^2$ of the virtual photon 
and the virtual photon energy fraction $z$ carried by the hadron.  
After integrating over the azimuthal angles, four variables are needed to describe the kinematics of the measured hadron: 
two inclusive ($Q^2$, $x_{Bj}$) and two hadronic ($p_T$, $z$).  

\section{The COMPASS Experiment}
The COMPASS experiment has been set up at the M2 muon beam line of the CERN SPS \cite{Abbon:2007pq}.  
Polarized 160\, GeV muons with an intensity of $2\cdot 10^8 \mu/4.8$\,s spill 
and a polarization of 80\% are scattered off a polarized $^6$LiD target.   
The target consists of two cells of opposite polarization which was reversed every 8\, h.  
The unpolarized sample is therefore the combination of the data from the two cells.    
The COMPASS spectrometer is a large acceptance two-stage spectrometer 
which covers the kinematic range from quasi-real photo production to the DIS region.
Both stages use hadron calorimeters and absorber walls for muon identification.  
The data presented here were taken during the year of 2004.

\section{Hadron kinematic distributions}
The charged hadron identification is kept as as simple as possible.  
The particles coming out of the primary vertex are either identified as hadron or muons.  
From these hadrons, the selection requires that they create signals in the detector situated upstream and downstream of the first magnet.  
This ensures that the track momentum and charge are well defined by the bending of the magnetic field.  
%For this first analysis, 
The COMPASS ability to identify hadrons, with a RICH detector, was not used, but is intended for further analysis.\\
In order to correct for event losses caused by the non-uniform acceptance of the COMPASS spectrometer, 
a full Monte Carlo (MC) simulation has been performed.  
The events were generated with LEPTO, transported through the COMPASS detector simulation program COMGEANT 
and the reconstruction software CORAL.  
From this MC sample, 4-dimensional acceptance tables have been determined.  
Although very similar, positive and negative hadrons have different tables.  
The systematic error has been estimated to 5\%.  Only statistical errors are shown in the figures.  
In this analysis, the hadrons are separated into 23 intervals in $Q^2$ (from 1 to 10\, GeV/c$^2$) and $x_{Bj}$ (from 0.004 to 0.12)
further subdivided into 8 intervals in $z$ (from 0.2 to 0.8).

\section{Results}
\begin{figure}
      \begin{center}
            \subfigure[EMC]{\epsfxsize=0.43\columnwidth \epsfbox{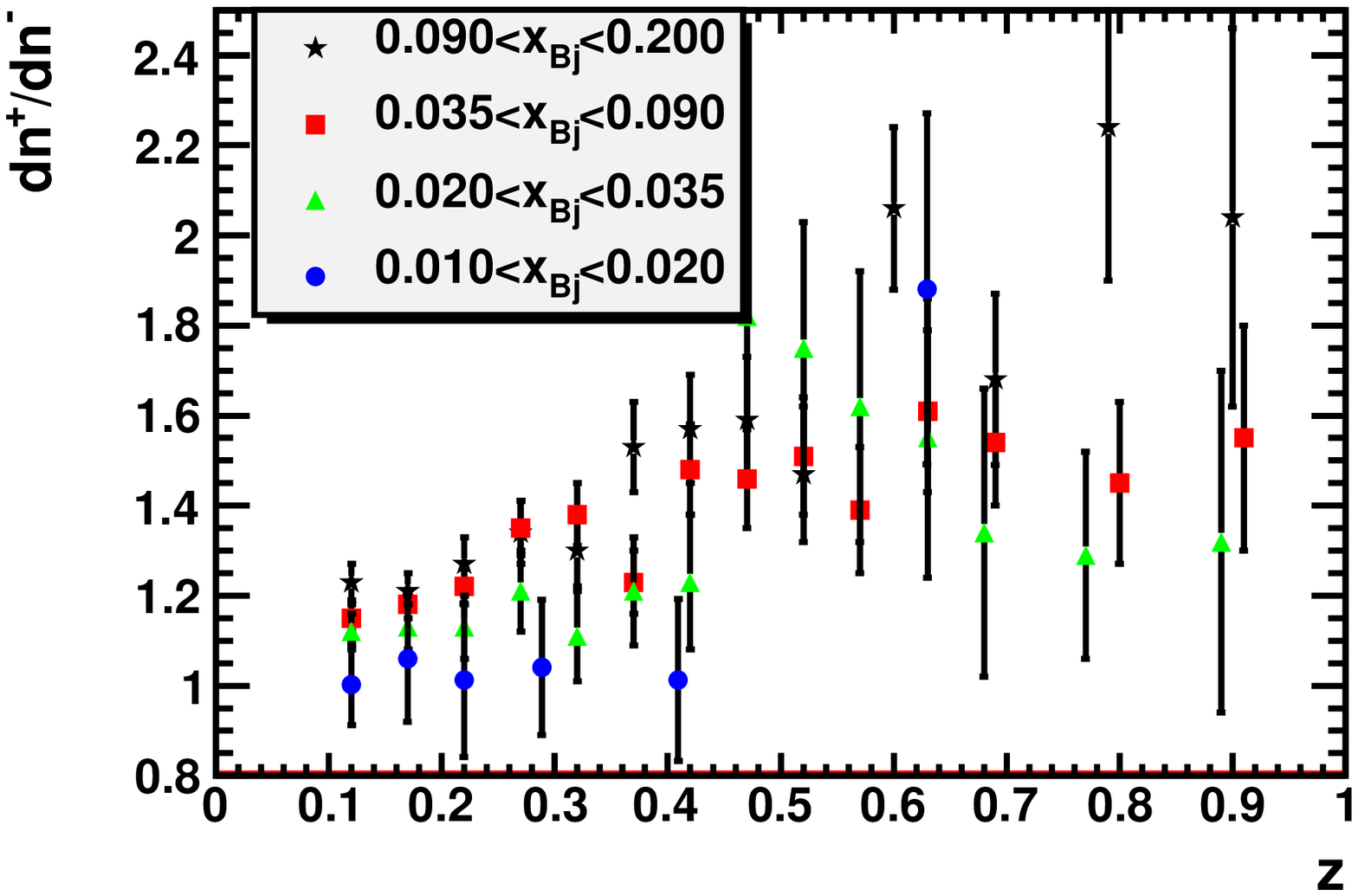}}
            \subfigure[COMPASS]{\epsfxsize=0.43\columnwidth \epsfbox{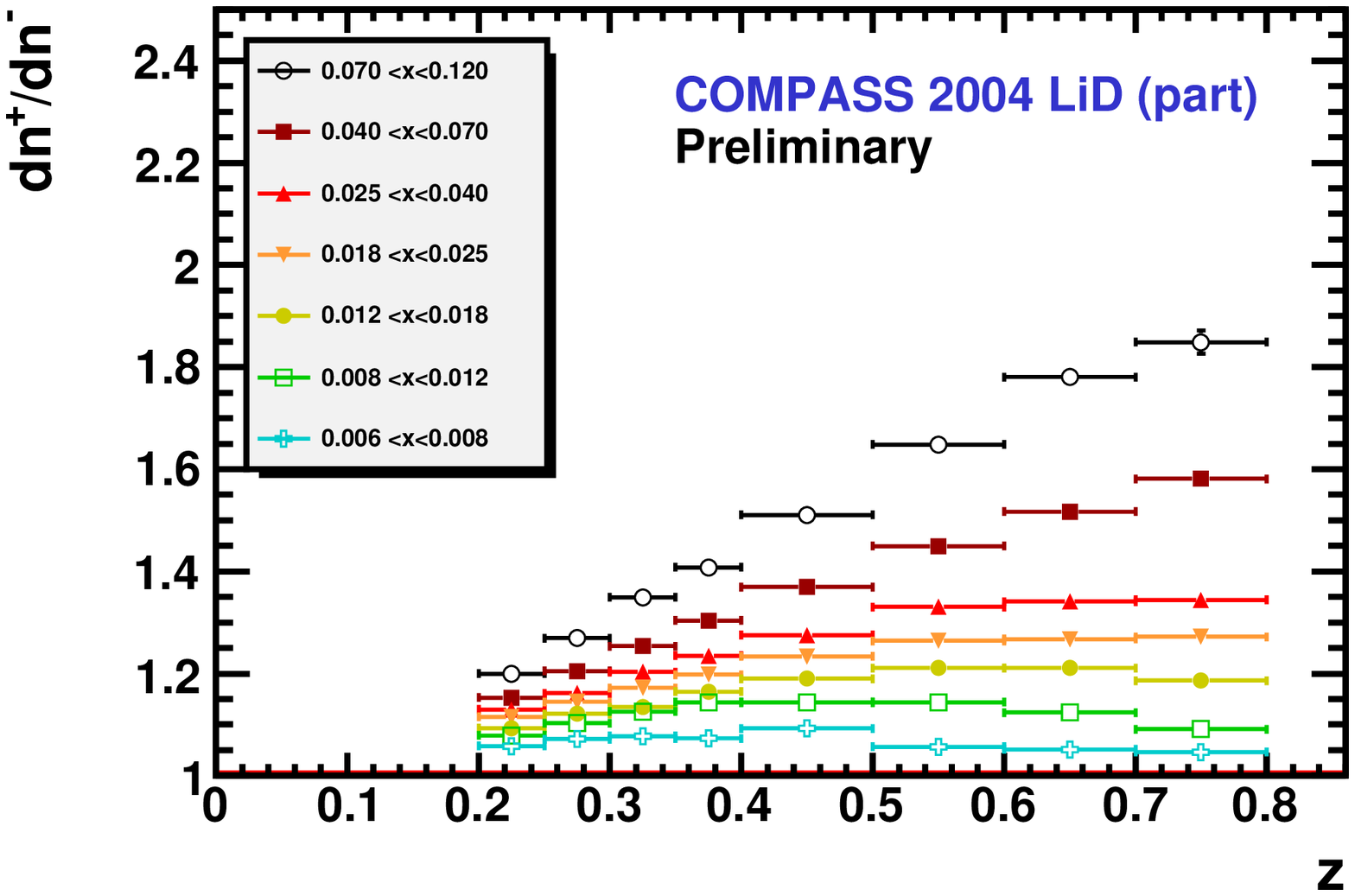}}
      \end{center}
\caption{Charged hadron multiplicity ratios $dn^+/dn^-$ as function of $z$ for EMC \cite{Ashman:1991cj} and COMPASS.}
\label{f_chgRatio}
\end{figure}

Hadron muoproduction has been studied for many years and the EMC experiment \cite{Ashman:1991cj} covered a similar kinematic range as COMPASS.  
An interesting comparison with previous data is the ratio of positive and negative hadrons because the acceptance is canceled to a good approximation.  
The hadron multiplicity ratios are shown in figure \ref{f_chgRatio}.  
COMPASS results show clearly the $z$ and $x_{Bj}$-dependence, 
where the fraction of positive hadrons increases with $x_{Bj}$ (getting closer to the valence region) and $z$ (more related to the struck parton).  
This agrees with the model of valence quarks where the positive quarks have a higher electric charge.\\

\subsection{$p_T^2$ distributions}
\begin{figure}
  \begin{center}
    \subfigure{\epsfxsize=0.43\columnwidth \epsfbox{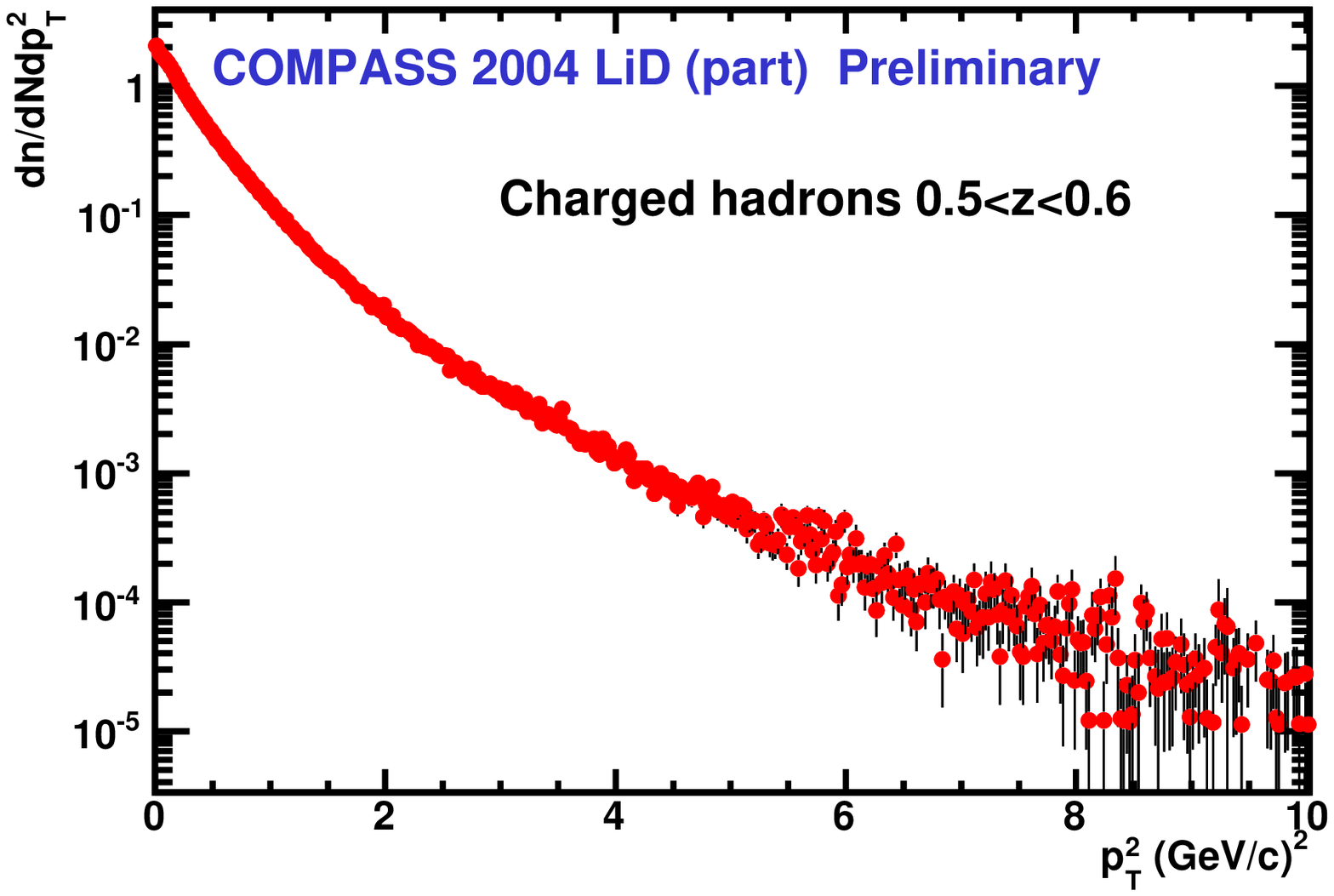}}
    \subfigure{\epsfxsize=0.43\columnwidth \epsfbox{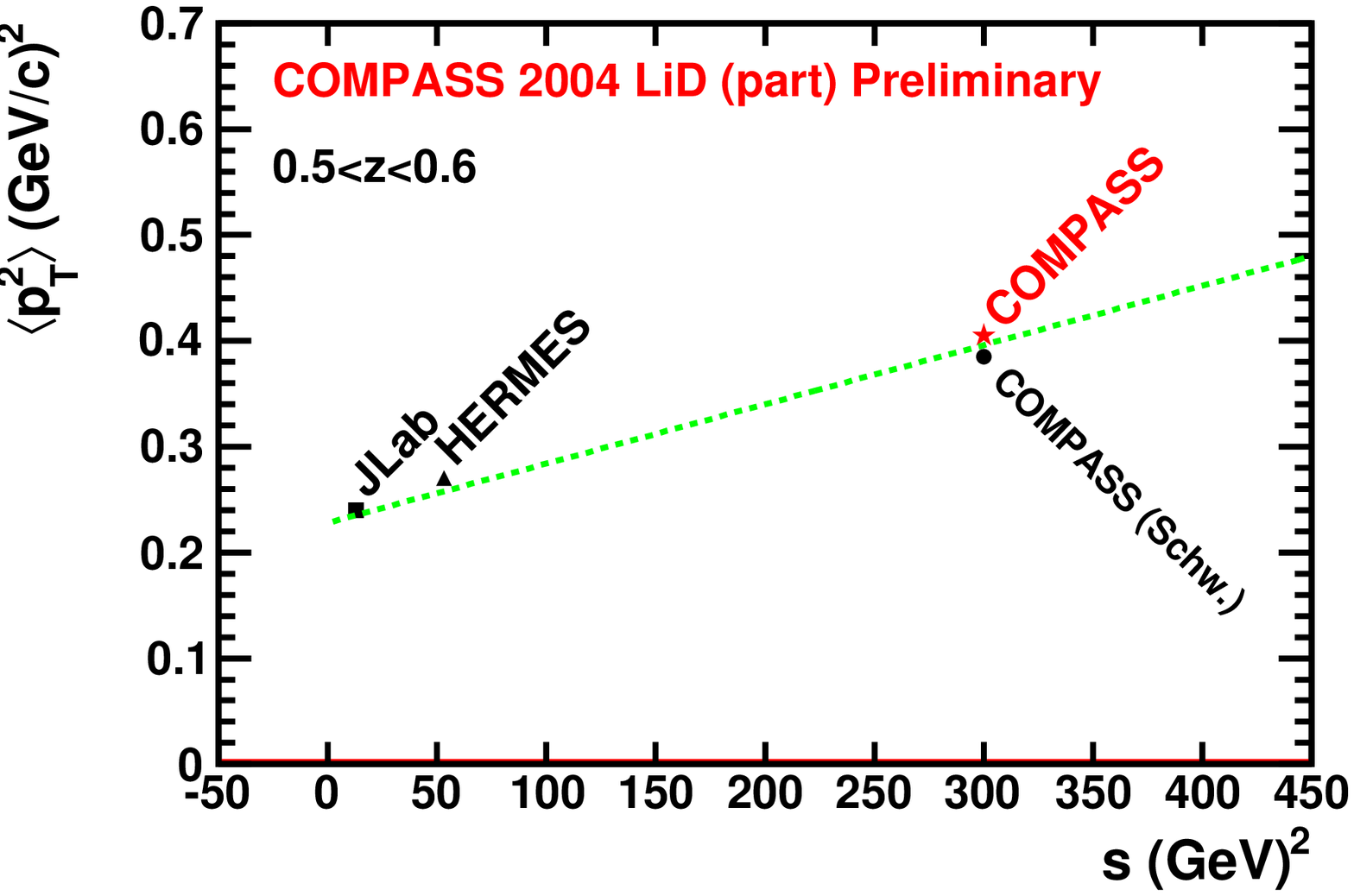}}
%\resizebox{0.75\columnwidth}{!}{
%\includegraphics{c_avpt2VSs_Schwt_corr.eps} }
  \end{center}
\caption{ The left figure shows the differential $p_T^2$ distribution of hadrons with $0.5<z<0.6$.  
It is used to determine the corrected COMPASS averaged over all $p_T^2$.  
This value is to correct the figure on the right from \cite{Schweitzer:2010tt}, where three experiments show the $s$-dependence of the average $p_T^2$.
The red star COMPASS point is the value from this analysis, the black dot, COMPASS (Schw.), is the value inferred by \cite{Schweitzer:2010tt} from data uncorrected for acceptance.}
\label{f_metz}
\end{figure}

\begin{figure}
  \begin{center}
    \subfigure{\epsfxsize=0.43\columnwidth \epsfbox{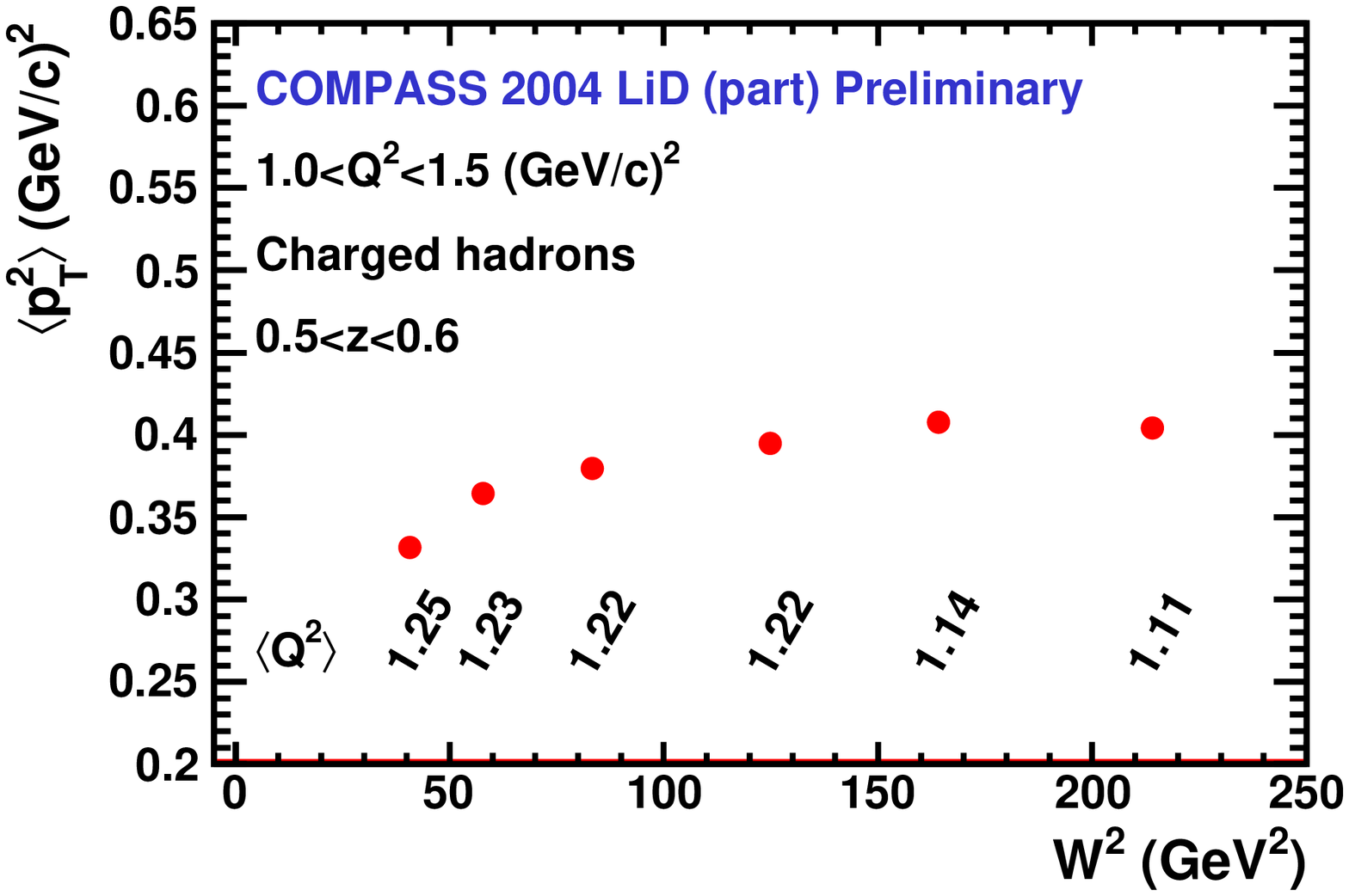}}
    %\subfigure{\epsfxsize=0.43\columnwidth \epsfbox{avpt2VSw2_Q1p5TO2p5.eps}}
    %\subfigure{\epsfxsize=0.43\columnwidth \epsfbox{avpt2VSw2_Q2p5TO3p5.eps}}
    \subfigure{\epsfxsize=0.43\columnwidth \epsfbox{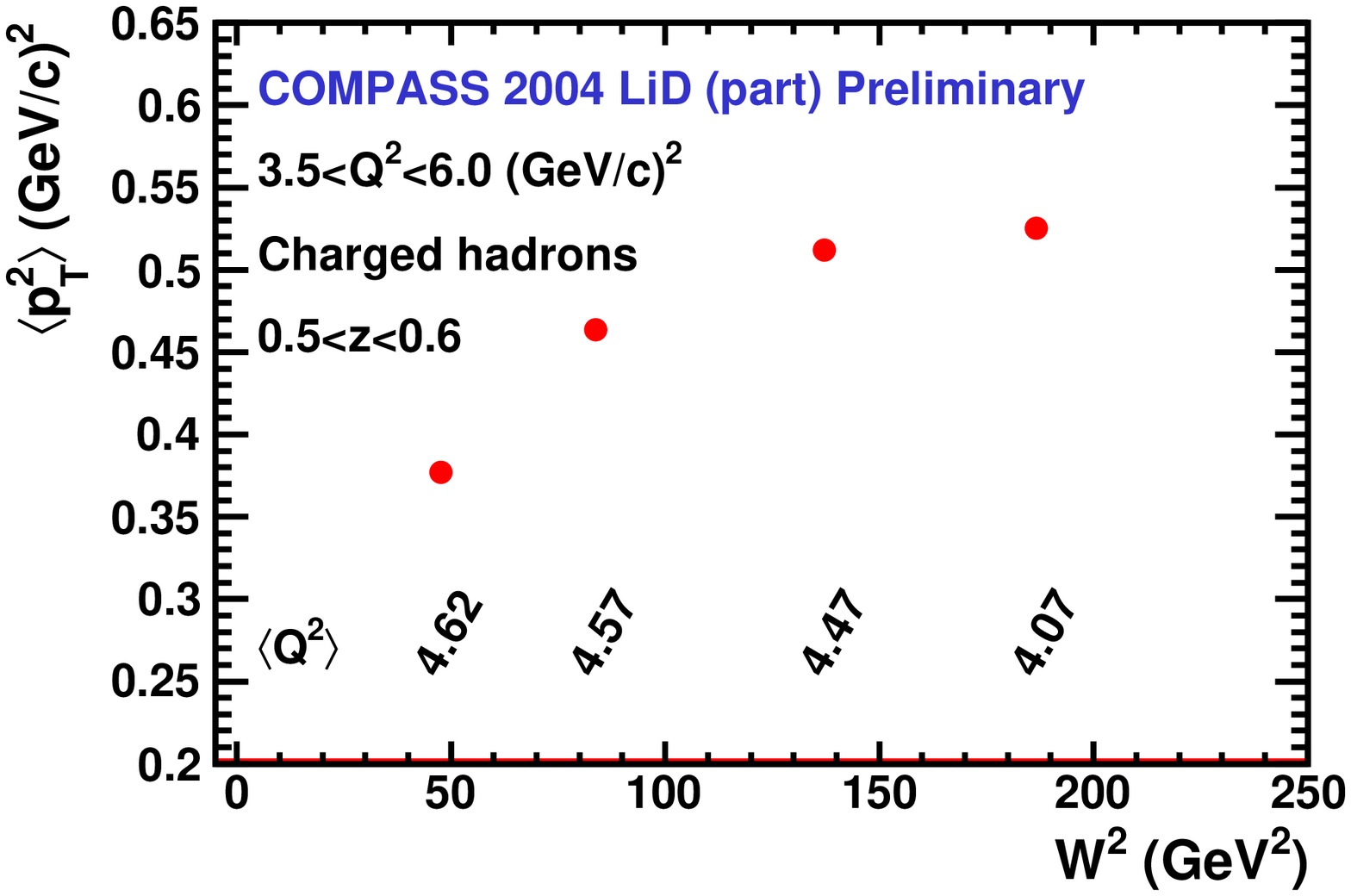}}
  \end{center}
\caption{Charged hadron average over all $p_T^2$ vs $W^2$.  Note that contrary to all the following figures of this article, $\langle p_T^2 \rangle$ defines a standard average over all $p_T^2$.}
\label{f_avptVSw2}
\end{figure}

%Before interpreting the results, the distributions can be compared with the previous EMC \cite{Ashman:1991cj} experiment covering kinematics of COMPASS.  
%An interesting comparison is the ratio of positive and negative hadrons because the acceptance is canceled to a good approximation.  
%The hadron ratio is shown in figure \ref{f_chgRatio}.\\
%\section{Energy dependence of the average $p_T^2$}
%\label{s_Edep}
%\subsection{Energy dependence of the average $p_T^2$}
%\label{ss_Edep}
According to \cite{Schweitzer:2010tt}, the average over all $p_T^2$ should depend linearly on the center of mass energy squared, $s$.  
They have verified their prediction with results from three fixed target experiments: JLab, HERMES and COMPASS.  
The value used for COMPASS was not corrected for acceptance, the value is corrected here as shown in figure \ref{f_metz}.  
The authors of \cite{Schweitzer:2010tt} note that the average $p_T^2$ should depend linearly on $W^2$ rather than $s$.  
The dependence is shown in figure \ref{f_avptVSw2} which is more compatible with a linear dependence on $W$ 
than on $W^2$.  The relation is not well established and, as mentioned in \cite{Schweitzer:2010tt}, 
the linear dependence on $s$ for Drell-Yan which inspired their SIDIS prediction, could also be a linear dependence on $\sqrt{s}$.  
Figure \ref{f_avptVSw2} suggests the latter is more accurate.
%\begin{figure}
%      \begin{center}
%            \subfigure[EMC]{\epsfxsize=0.43\columnwidth \epsfbox{c_charge_ratio_EMC.eps}}
%            \subfigure[COMPASS]{\epsfxsize=0.43\columnwidth \epsfbox{c_chgRatioVSz.eps}}
%      \end{center}
%\caption{Charged hadron differential distribution ratios $dn^+/dn^-$ as function of $z$ for EMC \cite{Ashman:1991cj} and COMPASS.}
%\label{f_chgRatio}
%\end{figure}

\subsection{Gaussian fit of the $p_T^2$ distributions and intrinsic transverse momentum}
\label{s_GaussFit}
\begin{figure}
  \begin{center}
    \subfigure{\epsfxsize=0.45\columnwidth \epsfbox{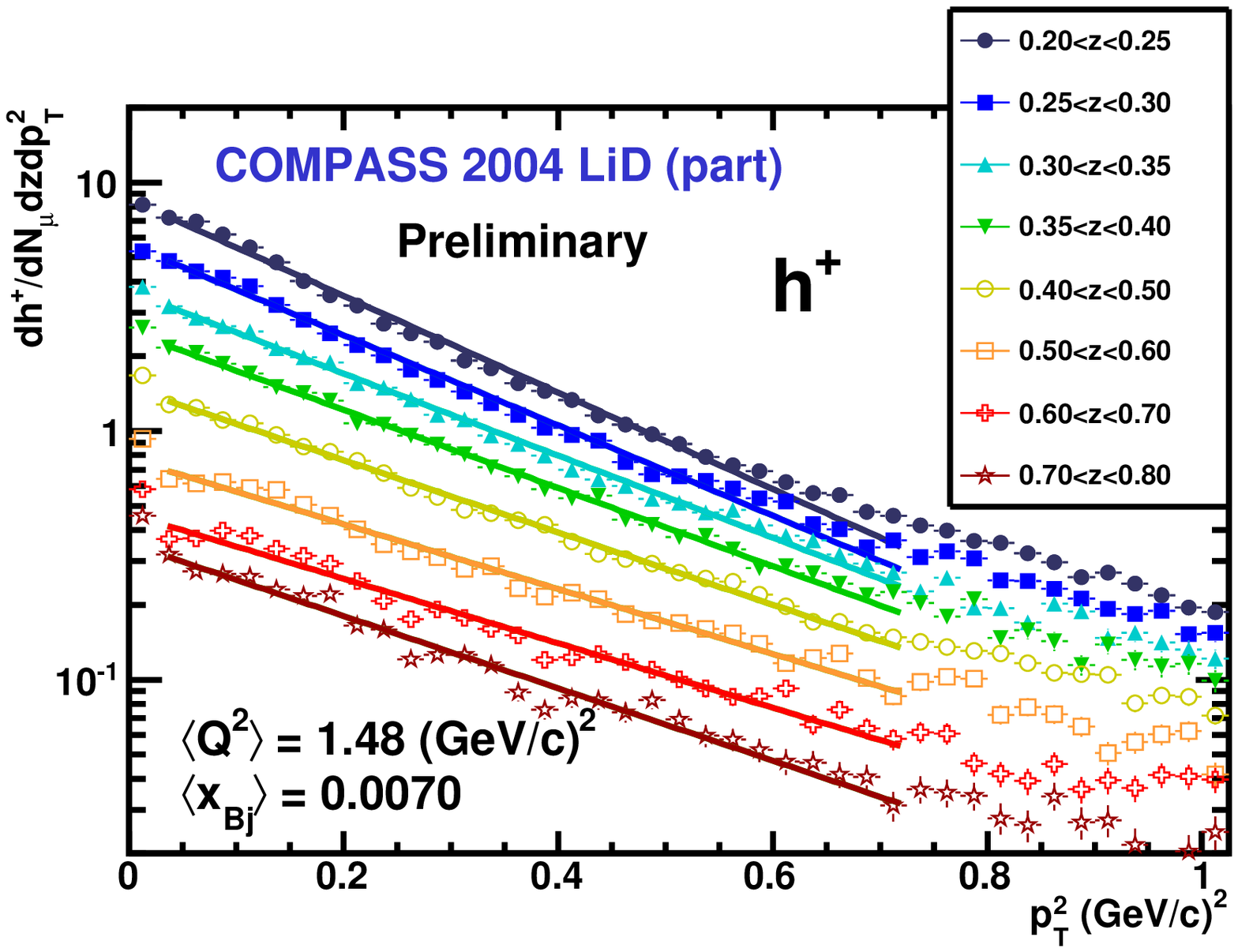}}
    \subfigure{\epsfxsize=0.45\columnwidth  \epsfbox{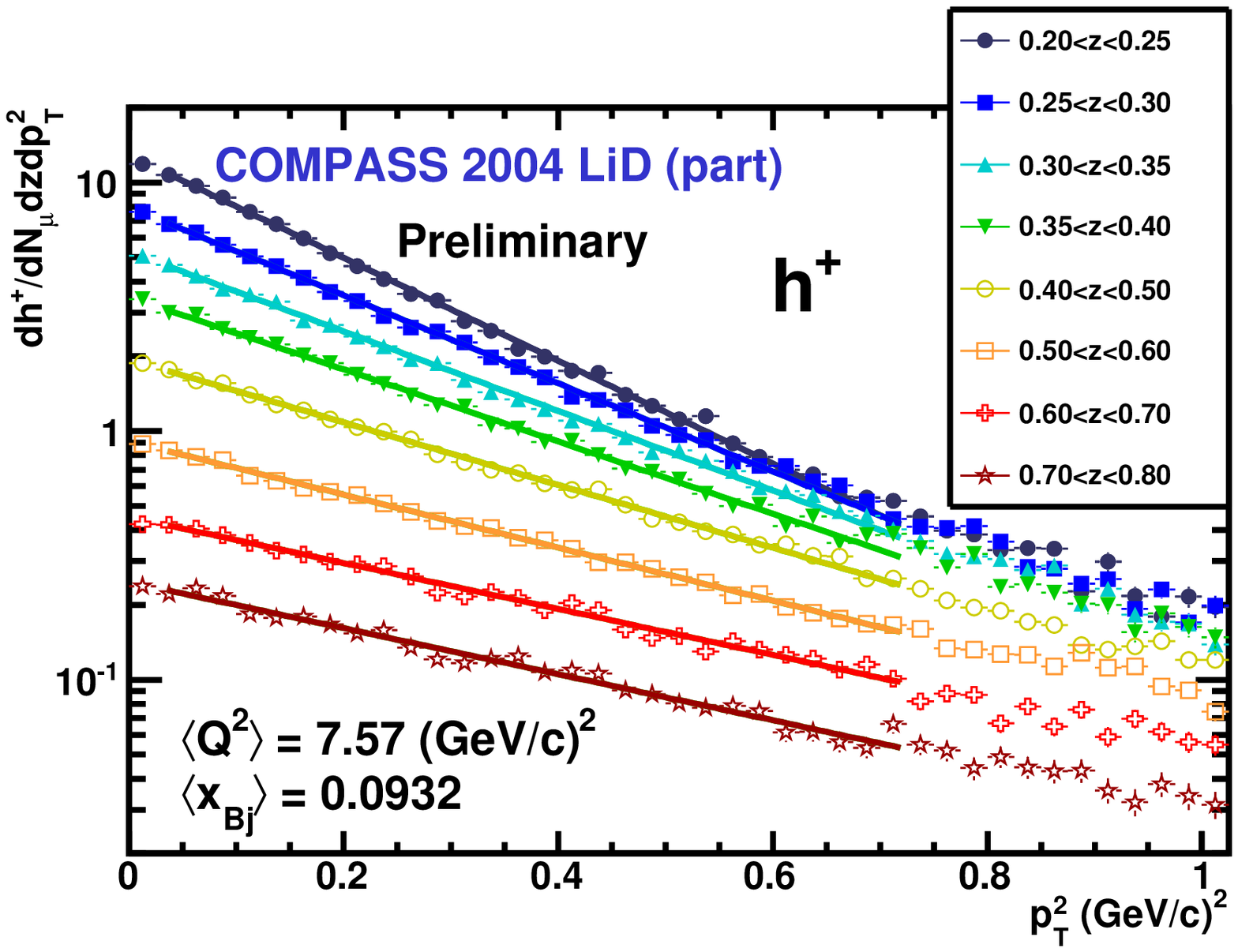}}
 \end{center}
 \caption{$p_T^2$ distributions fitted by a Gaussian for 
 ($1<Q^2<1.5$\, (GeV/c)$^2$, $0.006<x_{Bj}<0.008$) and ($6<Q^2<10$\, (GeV/c)$^2$, $0.07<x_{Bj}<0.12$) 
 subdivided into 8 $z$ intervals.}
\label{f_multNfit}
\end{figure}

\begin{figure}
  \begin{center}
    \subfigure{\epsfxsize=0.43\columnwidth \epsfbox{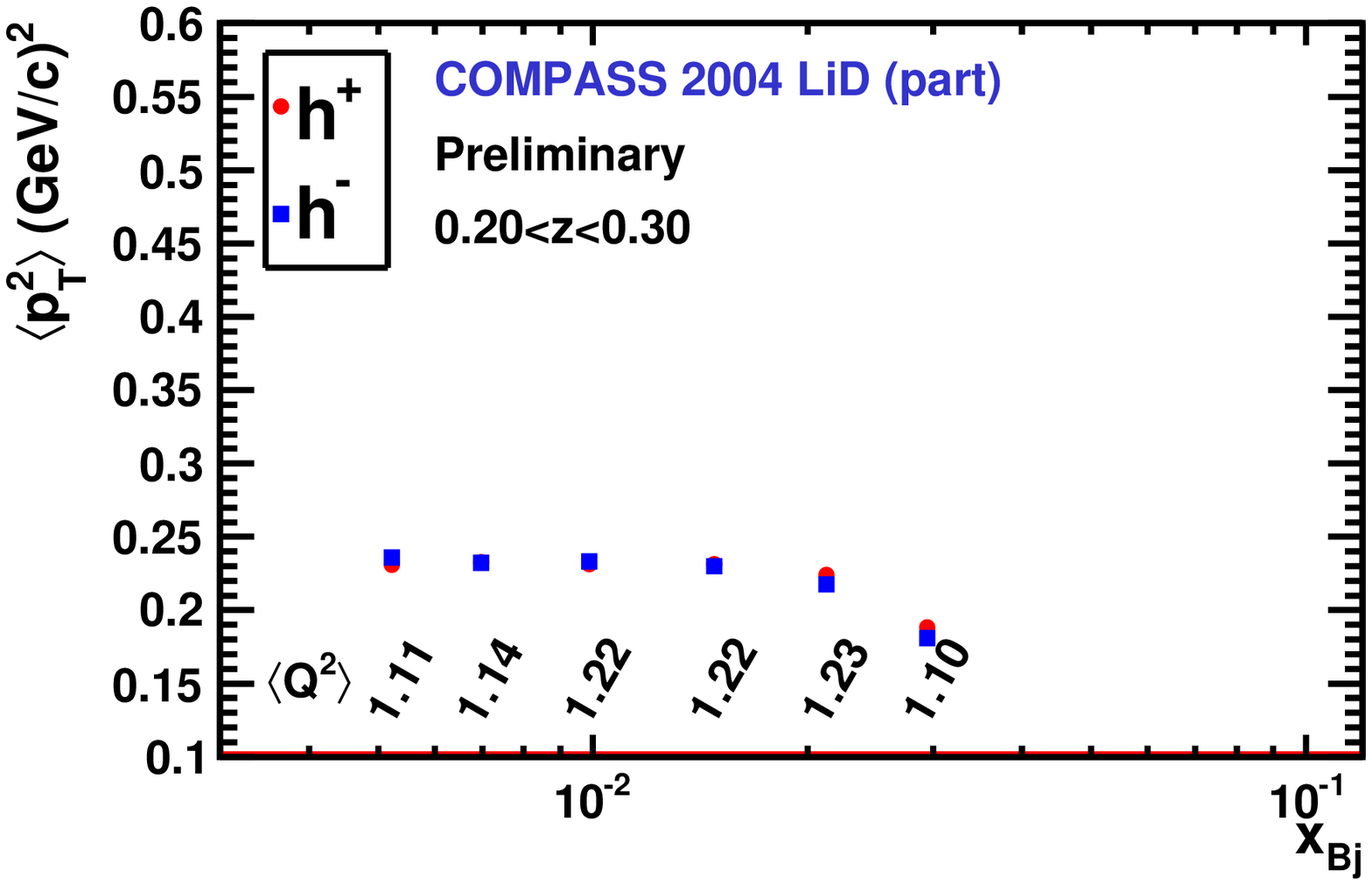}}
    \subfigure{\epsfxsize=0.43\columnwidth \epsfbox{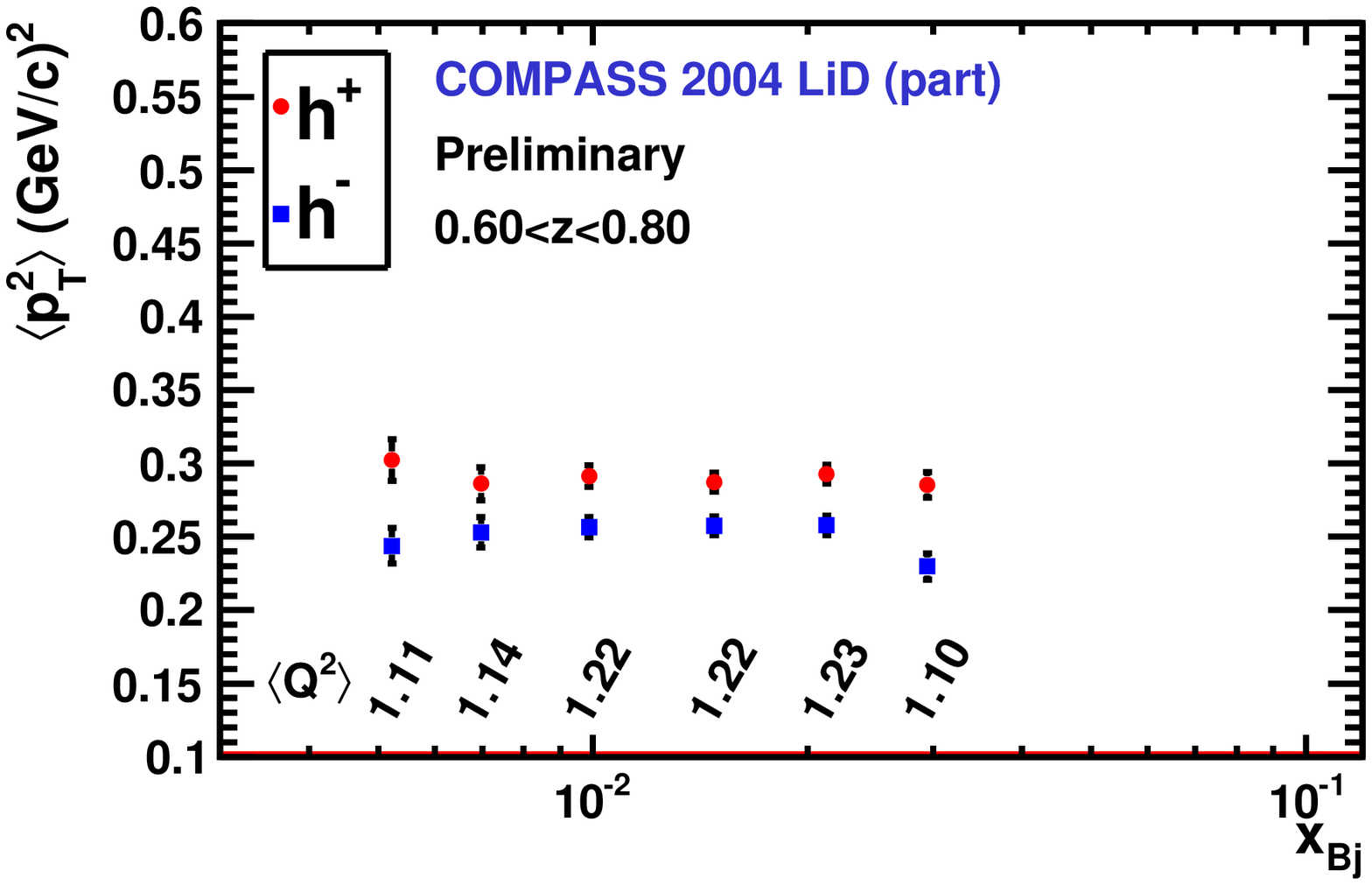}}\\
%    \subfigure{\epsfxsize=0.43\columnwidth \epsfbox{c_pt2VSx_41TO172_ALL.eps}}
%    \subfigure{\epsfxsize=0.43\columnwidth \epsfbox{c_pt2VSx_47TO178_ALL.eps}}
%    \subfigure{\epsfxsize=0.43\columnwidth \epsfbox{c_pt2VSx_71TO182_ALL.eps}}
%    \subfigure{\epsfxsize=0.43\columnwidth \epsfbox{c_pt2VSx_77TO188_ALL.eps}}\\
    \subfigure{\epsfxsize=0.43\columnwidth \epsfbox{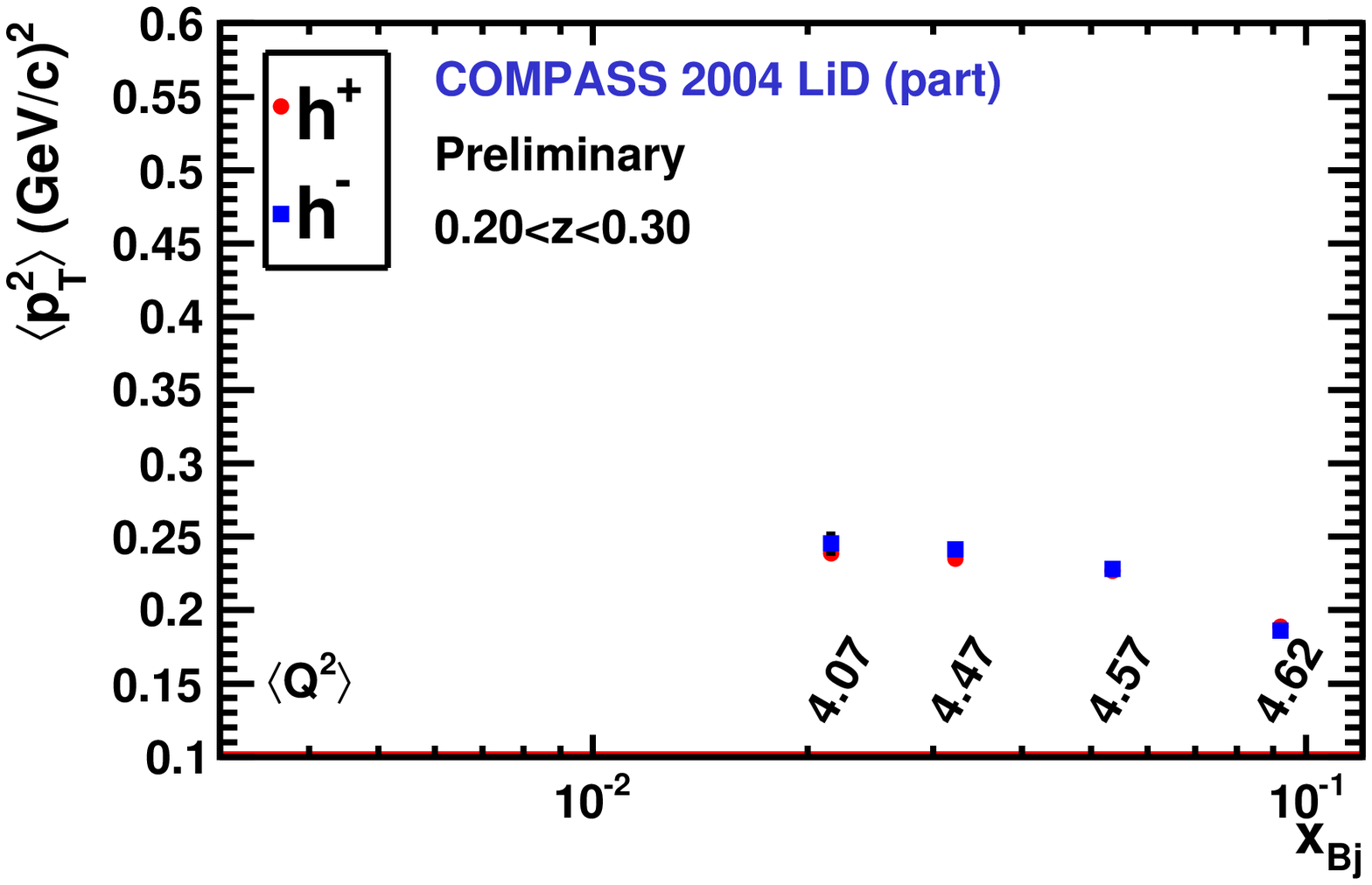}}
    \subfigure{\epsfxsize=0.43\columnwidth \epsfbox{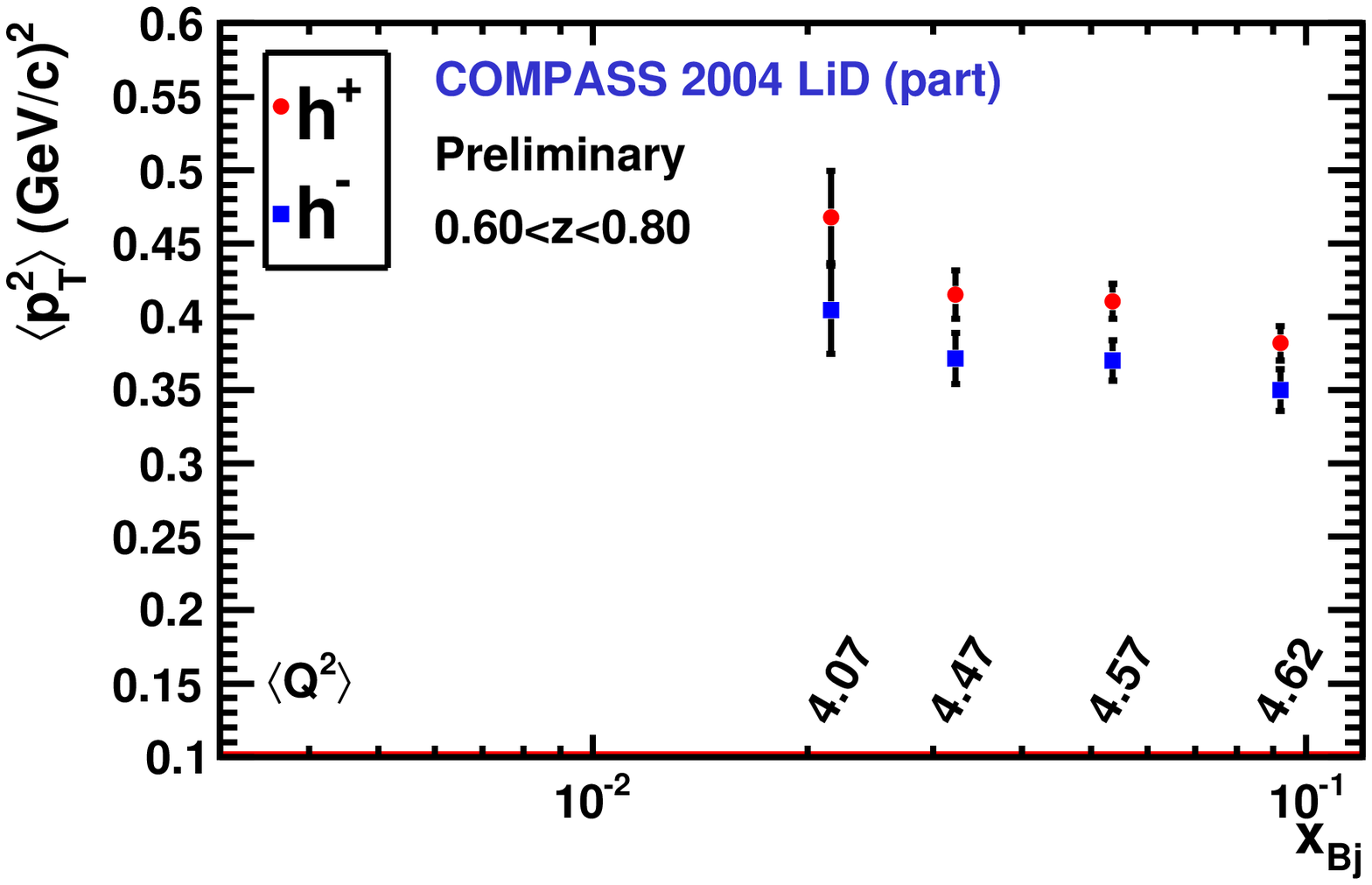}}
  \end{center}
\caption{Fitted $\langle p_T^2 \rangle$ vs $x_{Bj}$ for different $Q^2$ intervals for low (left column) and high (right column) $z$.  
In the present figure, the $\langle p_T^2 \rangle$ results from a fit over $p_T<0.85$\, GeV/c.}
\label{f_pt2VSx}
\end{figure}

\begin{figure}
  \begin{center}
    \subfigure{\epsfxsize=0.43\columnwidth \epsfbox{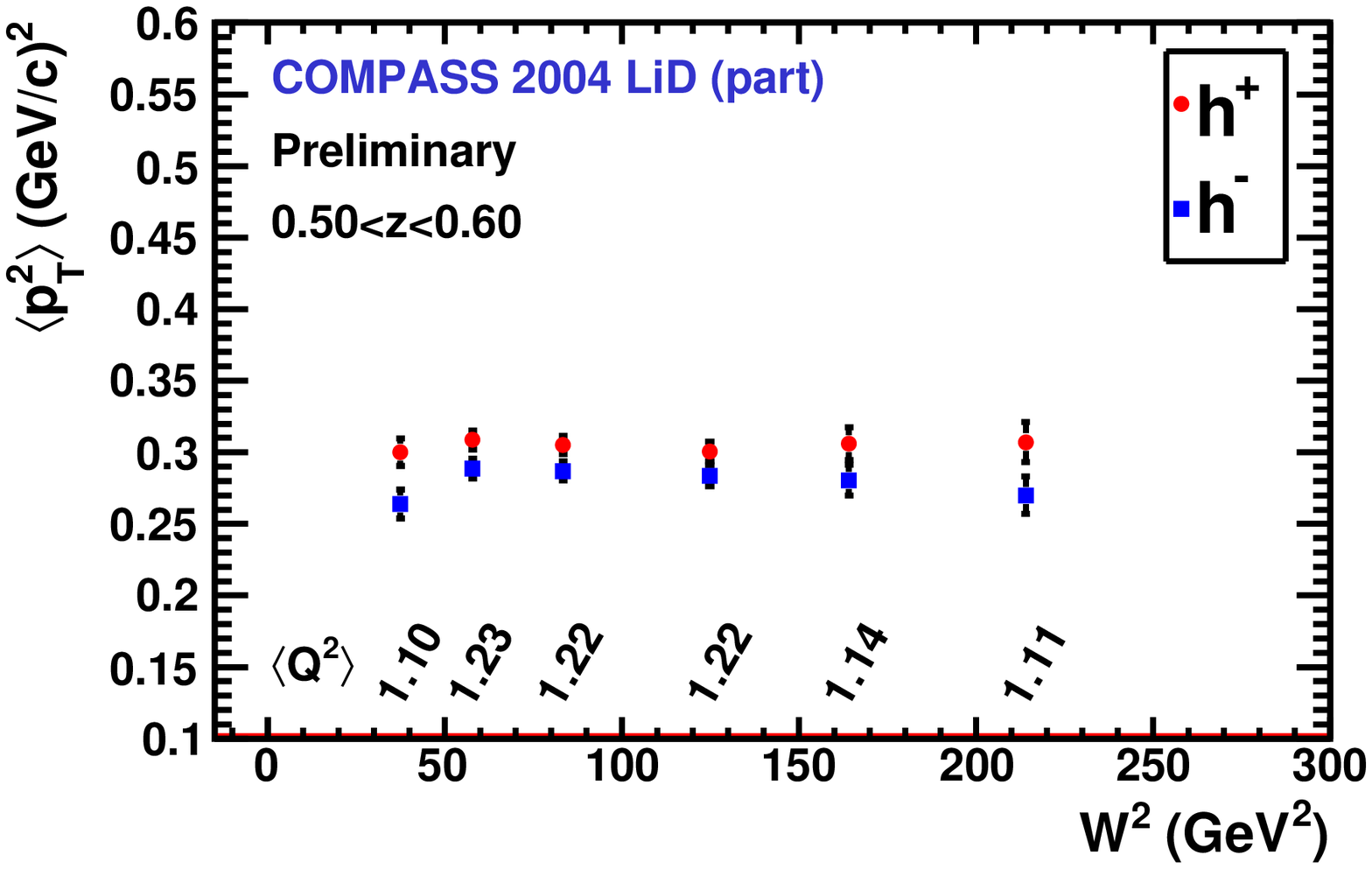}}
    %\subfigure{\epsfxsize=0.43\columnwidth \epsfbox{c_pt2VSw2_46TO176_ALL.eps}}\\
    %\subfigure{\epsfxsize=0.43\columnwidth \epsfbox{c_pt2VSw2_76TO186_ALL.eps}}
    \subfigure{\epsfxsize=0.43\columnwidth \epsfbox{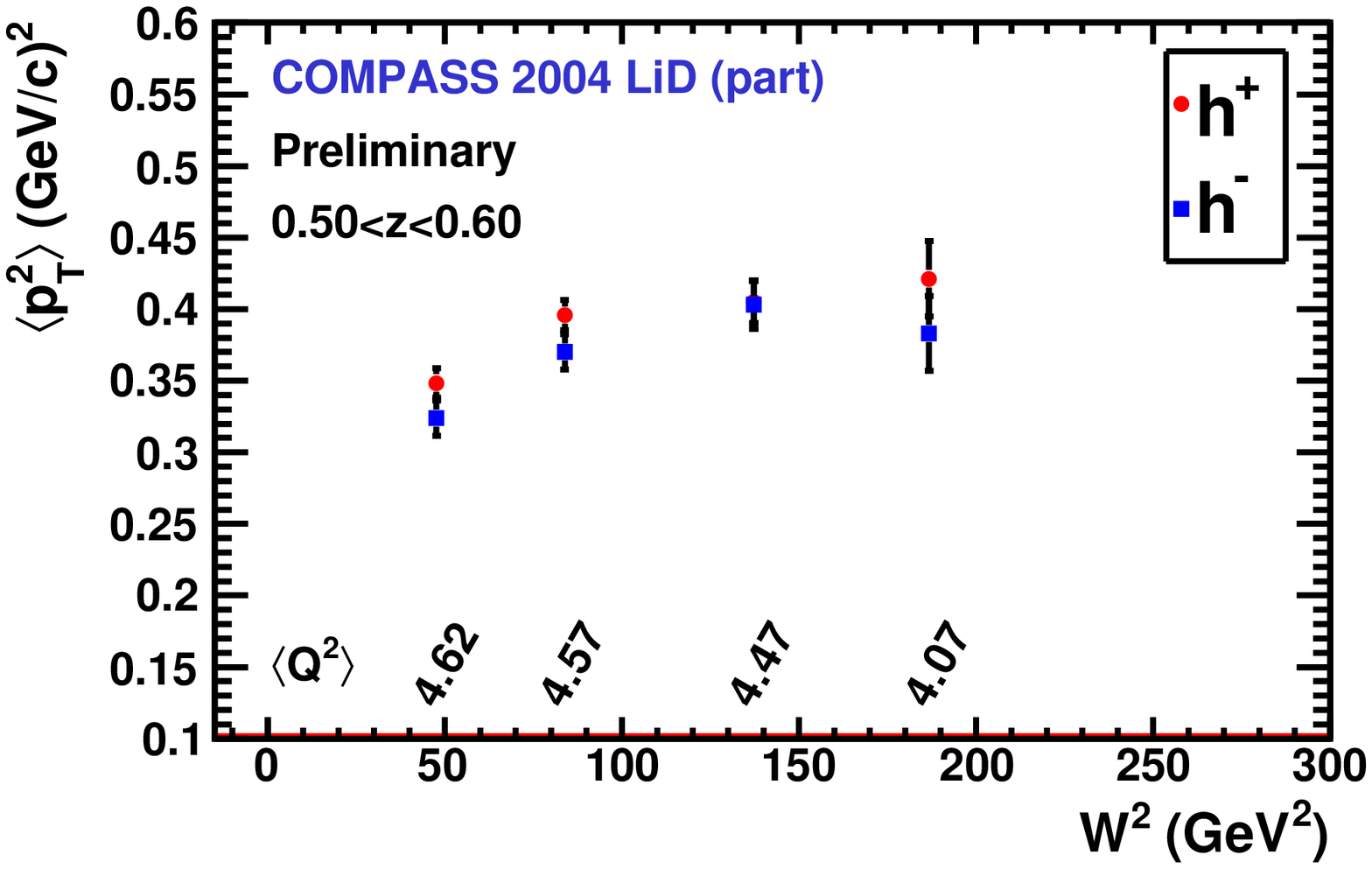}}
  \end{center}
\caption{Fitted $\langle p_T^2 \rangle$ vs $W^2$ at medium $z$ ([0.5, 0.6]) for different $Q^2$ intervals.  
This is to be compared with figure \ref{f_avptVSw2} where the average over all $p_T^2$ is plotted.  
In the present figure, the $\langle p_T^2 \rangle$ results from a fit over $p_T<0.85$\, GeV/c.}
\label{f_pt2VSw2}
\end{figure}

Using the Gaussian ansatz which assumes a Gaussian distribution of the intrinsic transverse momentum and of the 
transverse momentum acquired during fragmentation, the cross section is proportional to (cf. \cite{Anselmino:2005nn}):
\begin{equation}
      \frac{d^4\sigma^{\mu P \rightarrow \mu+h+X}}{dx_{Bj}dQ^2dzdp_T^2} 
      \propto \sum _q e_q^2
      f_q(x_{Bj}) D^h_q(z)
      \frac{e^{-p_T^2/\langle p_T^2 \rangle}}{\pi \langle p_T^2 \rangle},
\end{equation}
where 
\begin{equation}
\langle p_T^2 \rangle = \langle p_{\perp}^2 \rangle + z^2\langle k_{\perp}^2 \rangle .
\label{eq_pt2VSz2}
\end{equation}
The functions $f_q(x_{Bj})$ and $D^h_q(z)$ are the usual (integrated) distribution and fragmentation functions, respectively.  
The distributions and fits for two ($Q^2$, $x_{Bj}$) intervals and for all $z$ intervals are shown in figure \ref{f_multNfit}.
The fit is performed on the low $p_T$ interval [0.1, 0.85]\, GeV/c in order to stay away from pQCD effect.
The fitted $\langle p_T^2 \rangle$ as function of $x_{Bj}$ are shown in figure \ref{f_pt2VSx} for low and high $z$.  
A similar behavior was already observed by HERMES in \cite{Jgoun:2001ck} for the average $p_T$ (not from a fit but from a standard average).  
It is interesting to compare the average $p_T^2$ from the previous section (figure \ref{f_avptVSw2}) 
with the fitted $\langle p_T^2 \rangle$ for the middle $z$ shown in figure \ref{f_pt2VSw2}.
Contrary to the average aver all $p_T^2$, there is no clear $W^2$-dependence of the fitted $\langle p_T^2 \rangle$ 
(for 0.1$<p_T<$0.85 GeV/c) which is suppose to be unaffected by pQCD.\\

%\subsubsection{$z^2$-dependence and intrinsic transverse momentum extraction}
%\label{s_kt}

\begin{figure}
  \begin{center}
    \subfigure{\epsfxsize=0.43\columnwidth \epsfbox{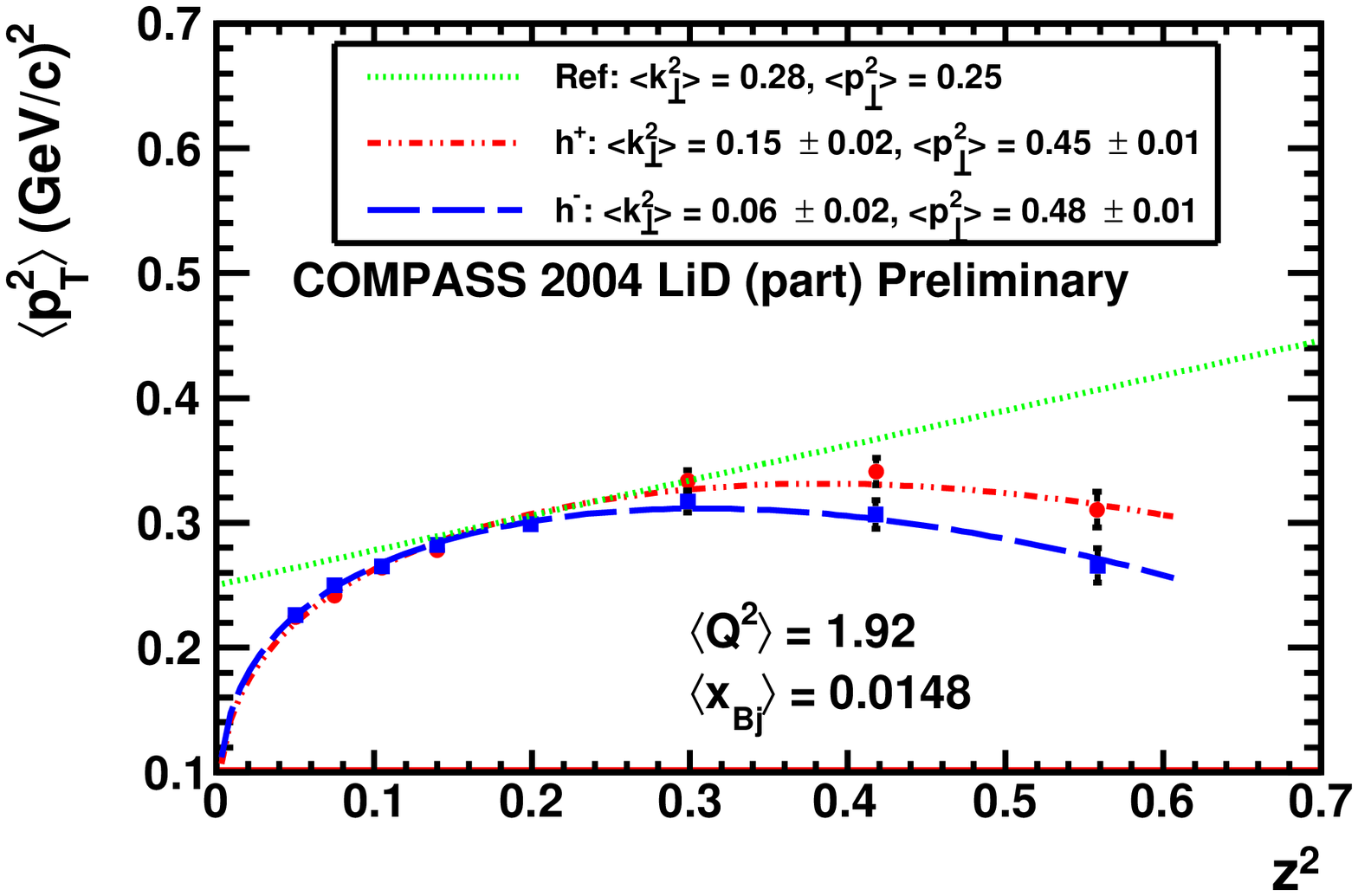}}
    \subfigure{\epsfxsize=0.43\columnwidth \epsfbox{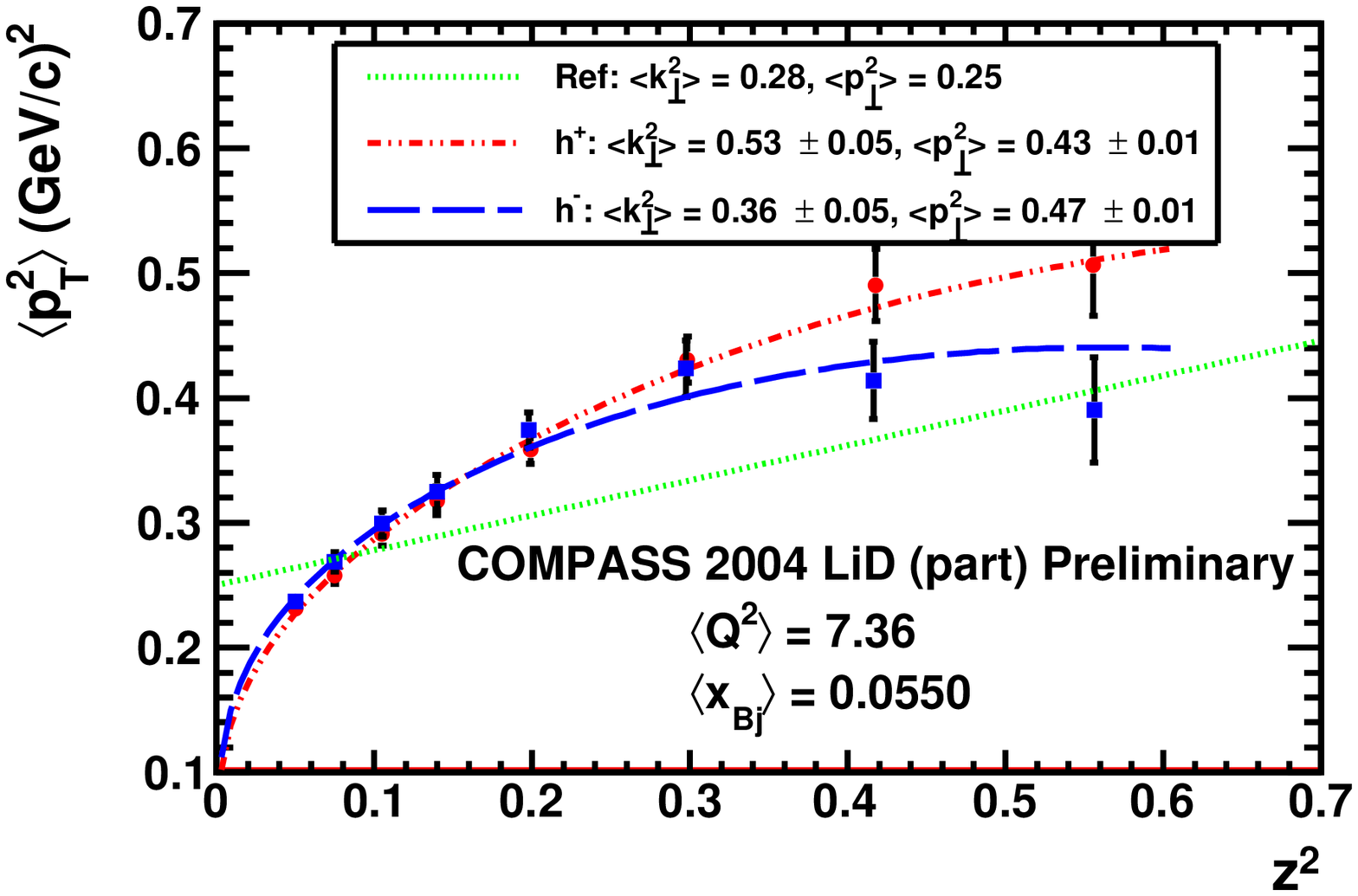}}
  \end{center}
\caption{Fitted $\langle p_T^2 \rangle$ vs $z^2$ for two ($Q^2$, $x_{Bj}$) intervals.  The fit function is given by equation (\ref{eq_relation_pt_pperp_kperp_frag}).
The dotted green line is the result of a fit, performed by \cite{Anselmino:2006rv}, to data from many experiments.  
In the present figure, the $\langle p_T^2 \rangle$ results from a fit over $p_T<0.85$\, GeV/c.}
\label{f_pt2VSz2}
\end{figure}

\begin{figure}
  \begin{center}
    \subfigure[$h^{+}$]{\epsfxsize=0.43\columnwidth \epsfbox{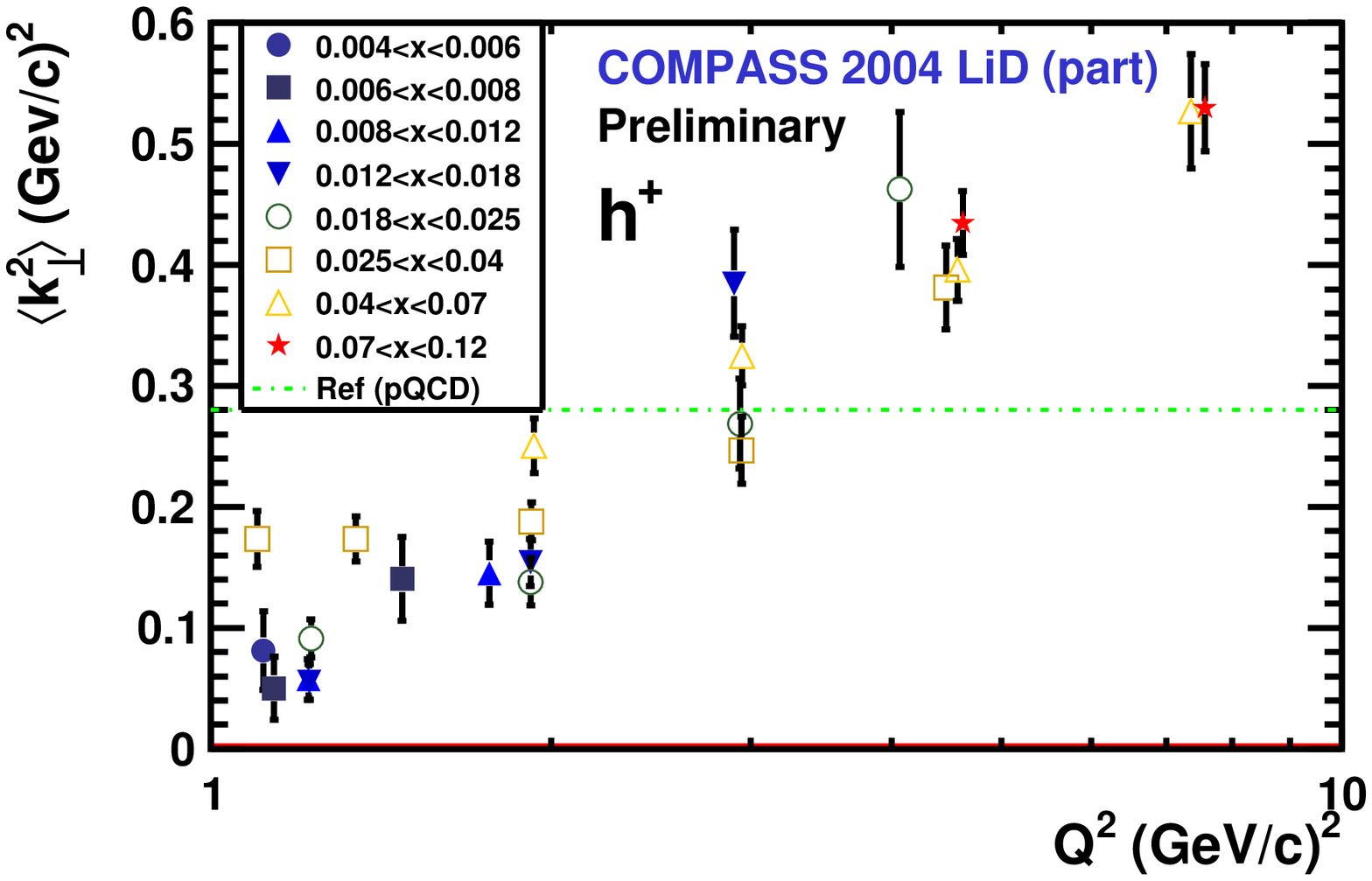}}
    \subfigure[$h^{-}$]{\epsfxsize=0.43\columnwidth \epsfbox{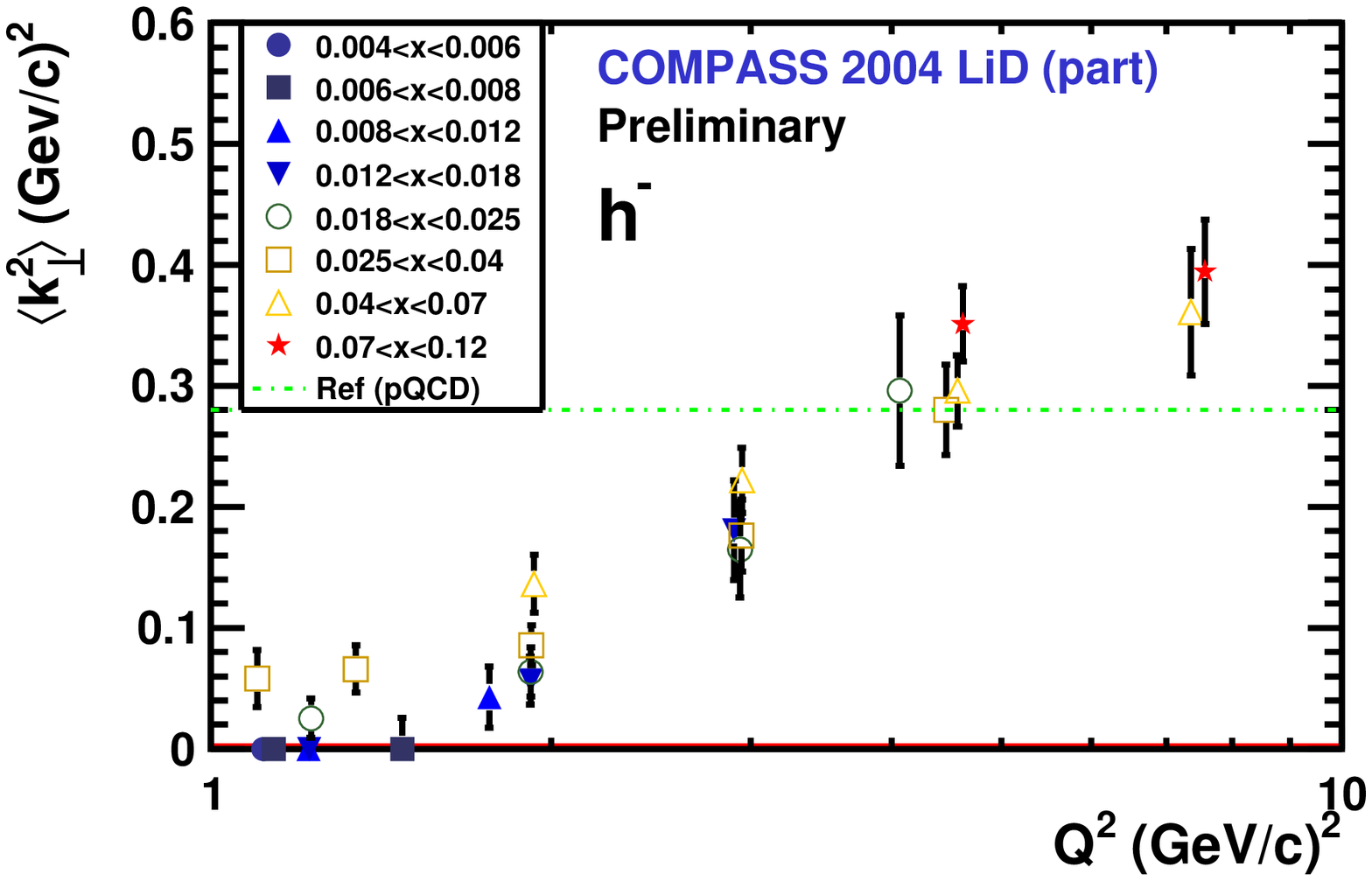}}
  \end{center}
\caption{Extracted $\langle k_{\perp}^2 \rangle$ vs $Q^2$ for various $x_{Bj}$ intervals. 
The dotted green line is the result of a fit, performed by \cite{Anselmino:2006rv}, to data from many experiments.}
\label{f_ktVSq2}
\end{figure}

\begin{figure}
  \begin{center}
    \subfigure{\epsfxsize=0.3\columnwidth \epsfbox{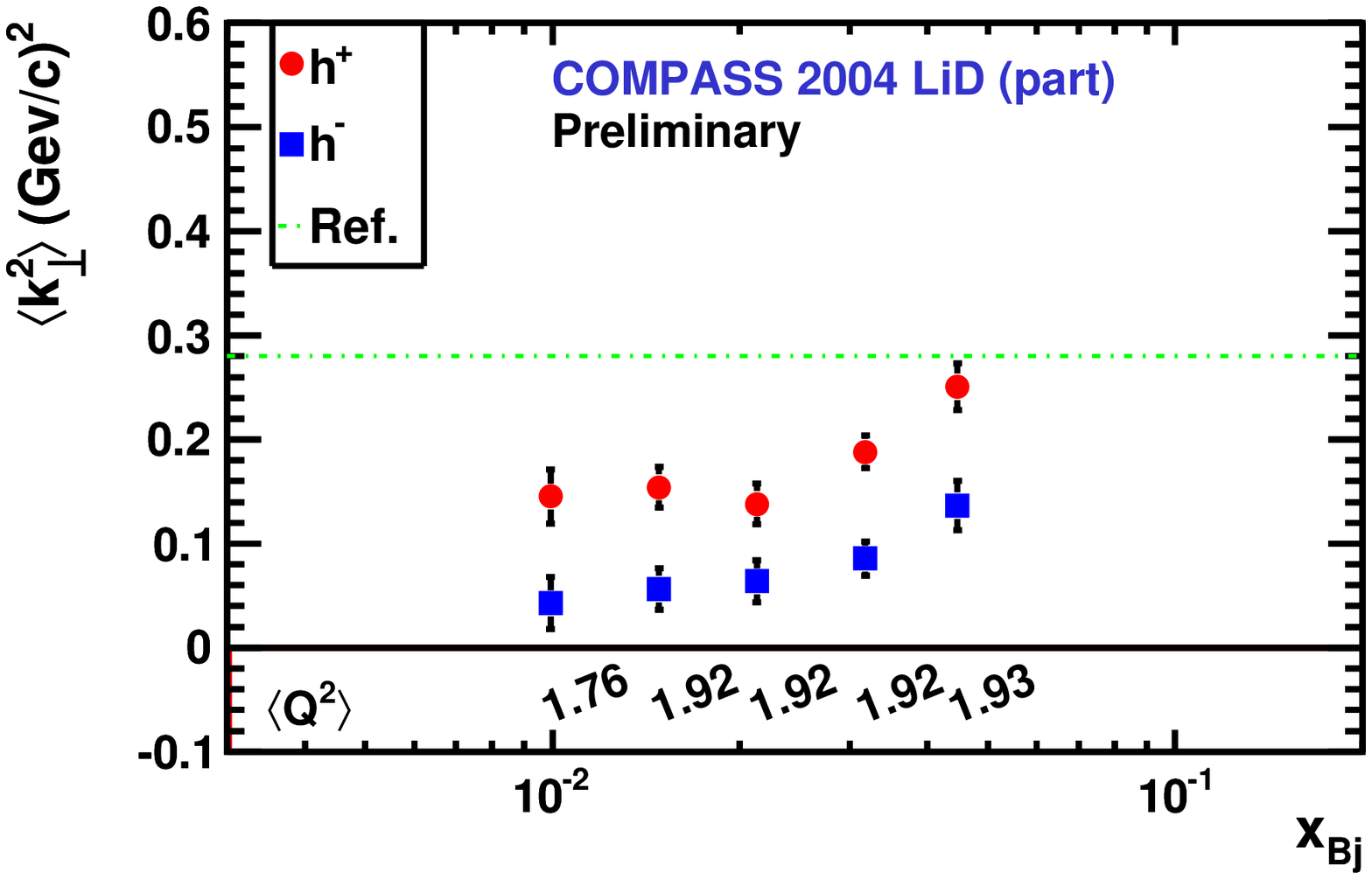}}
    \subfigure{\epsfxsize=0.3\columnwidth \epsfbox{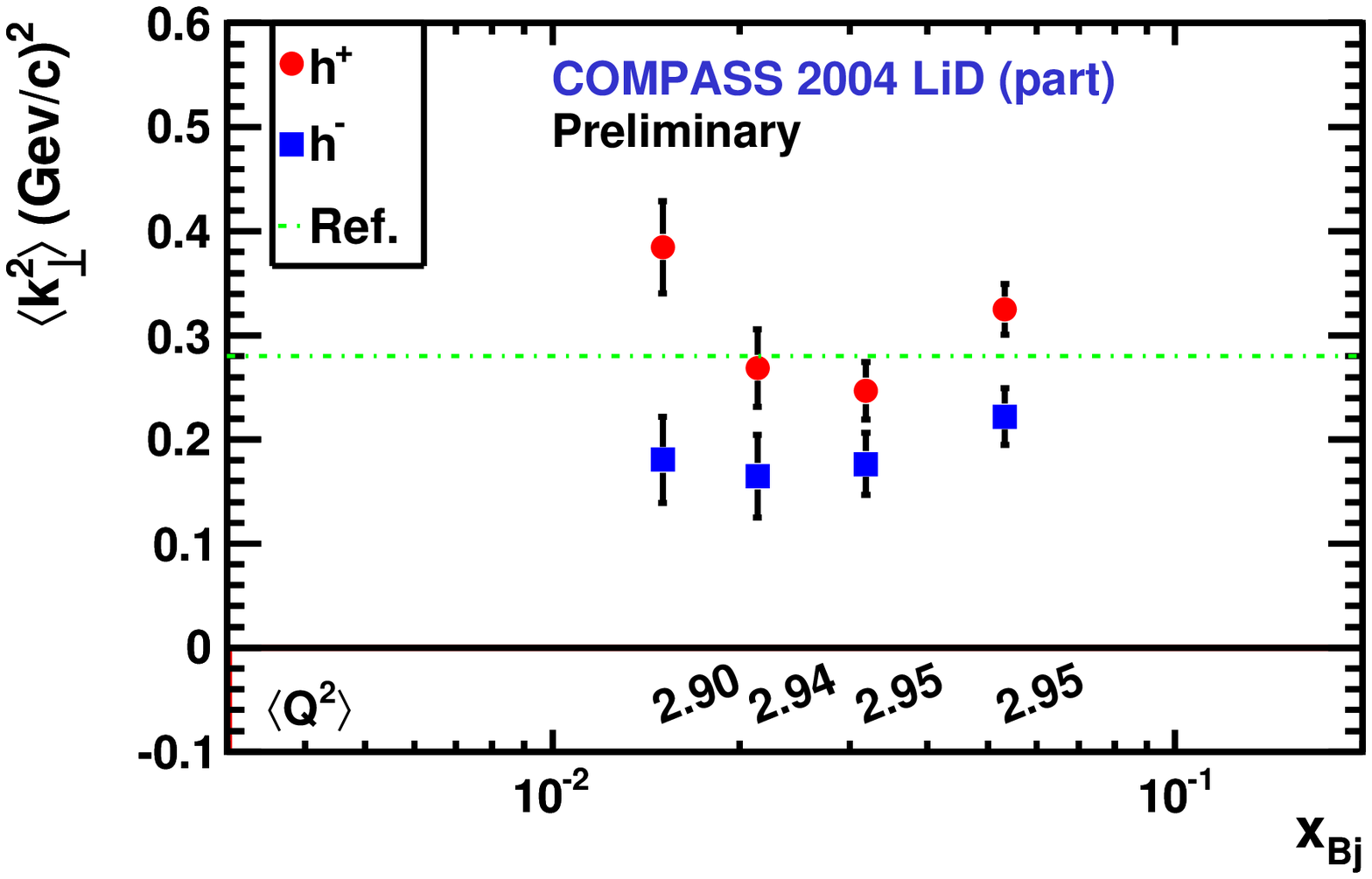}}
    \subfigure{\epsfxsize=0.3\columnwidth \epsfbox{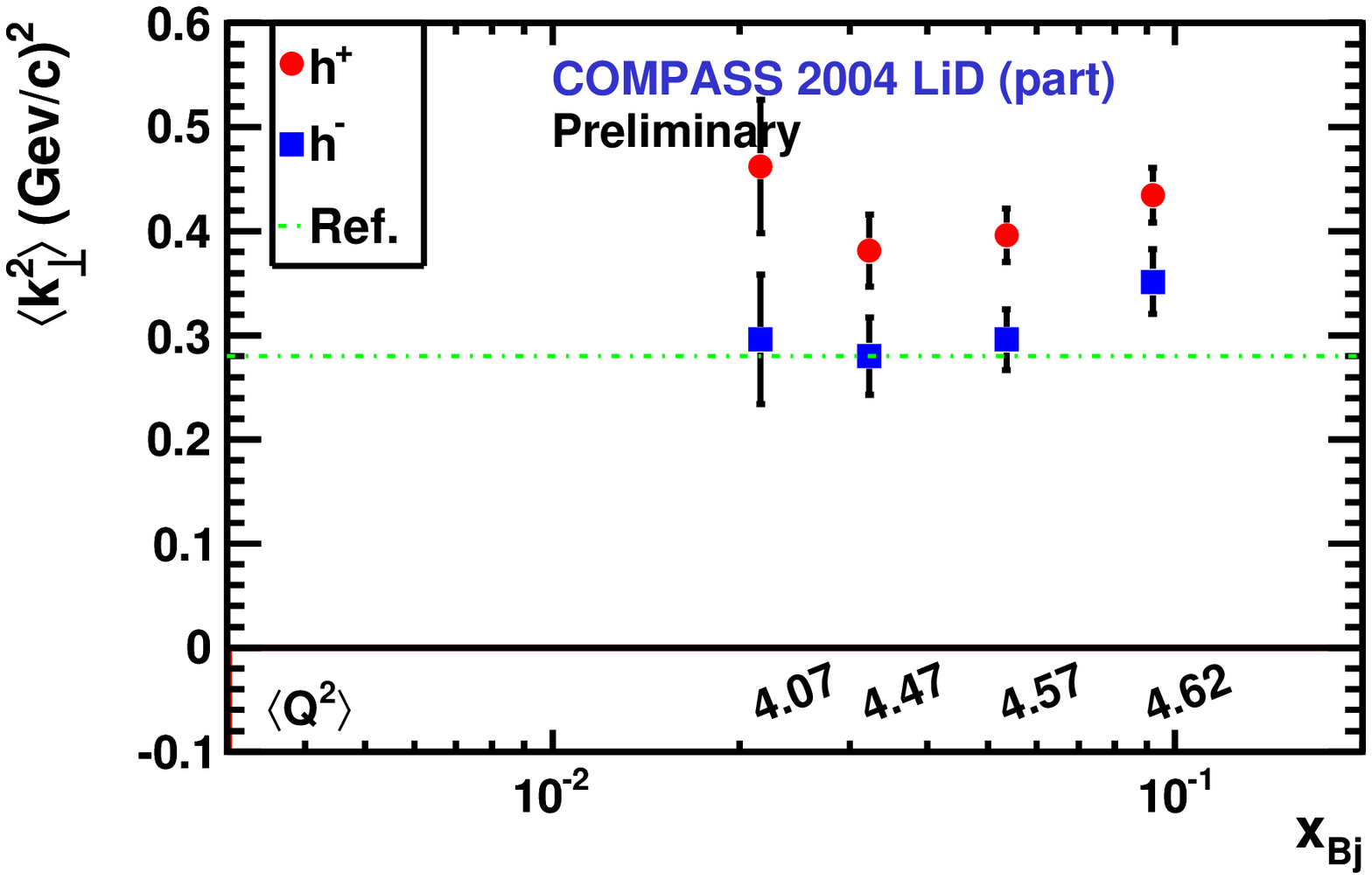}}
  \end{center}
\caption{Extracted $\langle k_{\perp}^2 \rangle$ vs $x_{Bj}$ for positive and negative hadron comparison. 
%The dotted green line is the fitted value on many experiments determined by \cite{Anselmino:2006rv}.}
The dotted green line is the result of a fit, performed by \cite{Anselmino:2006rv}, to data from many experiments.}
\label{f_ktVSx}
\end{figure}

The $z^2$-dependence of the fitted $\langle p_T^2 \rangle$ is of particular interest 
because of its relation to the intrinsic transverse momentum through equation (\ref{eq_pt2VSz2}).  
The fitted $\langle p_T^2 \rangle$ for different $z^2$ for two ($Q^2$, $x_{Bj}$) intervals are shown in figure \ref{f_pt2VSz2}.  
The relation between $\langle p_T^2 \rangle$ and $z^2$ is certainly not linear as in equation (\ref{eq_pt2VSz2}).   
If a $z$-dependence of the transverse momentum acquired during fragmentation, $p_{\perp}$, is added such that
\begin{equation}
\langle p_T^2 \rangle = z^{\alpha}(1 - z)^{\beta}\langle p_{\perp}^2 \rangle + z^2\langle k_{\perp}^2 \rangle ,
\label{eq_relation_pt_pperp_kperp_frag}
\end{equation}
where $\alpha=0.5$ and $\beta=1.5$, the relation can be nicely fitted as shown in figure \ref{f_pt2VSz2}.
From this fit, the intrinsic momentum, $\langle k_{\perp}^2 \rangle$, for various ($Q^2$, $x_{Bj}$) intervals can be extracted.  
The extracted $\langle k_{\perp}^2 \rangle$ as function of $Q^2$ are shown in figure \ref{f_ktVSq2} 
and as function of $x_{Bj}$, comparing results from positive and negative hadrons, are shown in \ref{f_ktVSx}.

\section{Conclusion}
\label{s_conclusion}
The differential $p_T^2$ distributions of charged hadrons produced by muons scattered off a $^6$LiD target have been determined for various kinematic intervals.  
The low $p_T^2$ have been fitted with a Gaussian at different $z$ 
such that the intrinsic transverse momentum could be extracted in the framework of the Gaussian ansatz.  
The non linear relation between the fitted $\langle p_T^2 \rangle$ and $z^2$ have been reproduced 
by adding a z-dependence of the transverse momentum acquired during fragmentation.  
The extracted $\langle k_{\perp}^2 \rangle$ shows a clear dependence on $Q^2$ and a possible dependence on $x_{Bj}$, 
although less conspicuous for higher $Q^2$, where the framework is more reliable.  
%The other obvious feature is a $\langle k_{\perp}^2 \rangle$ systematically higher for positive hadrons compared to negative hadrons.  
Also, $\langle k_{\perp}^2 \rangle$ is systematically higher for positive hadrons compared to negative hadrons.  
This suggests a flavor dependence of the intrinsic transverse momentum.  
%This analysis could be continued in many ways.  
%Most importantly, COMPASS ability to identify hadrons would improve purity of the relation with the structure of the nucleon;
This behavior could be further investigated using COMPASS ability to identify hadrons; 
kaon identification could provide access to characteristics of the strange quark TMDs.\\

\noindent \textbf{Acknowledgment}\\
\noindent I would like to thank Dr. Alessandro Bacchetta for his helpful comments.


\begin{thebibliography}{}
%% and use \bibitem to create references.
\bibitem{Abbon:2007pq}
Abbon, P. and others, Nucl. Instrum. Meth.\textbf{A577}, (2007) 455-518
\bibitem{Anselmino:2005nn}
%% Format for Journal Reference
Anselmino, M. and others, Phys. Rev. \textbf{D71}, (2005) 074006
\bibitem{Anselmino:2006rv}
%% Format for Journal Reference
Anselmino, M. and Boglione, M. and Prokudin, A. and Turk, C., Eur. Phys. J. \textbf{A31}, (2007) 373-381
\bibitem{Schweitzer:2010tt}
%% Format for Journal Reference
Schweitzer, P. and Teckentrup, T. and Metz, A., Phys. Rev. \textbf{D81}, (2010) 094019
\bibitem{Ashman:1991cj}
%% Format for Journal Reference
Ashman, J. and others, Z. Phys. \textbf{C52}, (1991) 361-388
%\bibitem{Cahn:1978se}
%Cahn, Robert N., Phys. Lett. \textbf{B78}, (1978) 269
\bibitem{Jgoun:2001ck}
%% Format for Journal Reference
Jgoun, Anton on behalf of the HERMES collaboration, Talk given at 36th Rencontres de Moriond on QCD and Hadronic Interactions, (2001)
%% Format for books
%\bibitem{RefB}
%Author, \textit{Book title} (Publisher, place year) page numbers
%% etc
\end{thebibliography}
\end{document}